\documentclass[aps,pre,twocolumn,amsmath,amssymb,showpacs,amsfonts]{revtex4}
\usepackage{epsfig}
\usepackage{graphicx}
\usepackage{dcolumn}
\usepackage{bm}

\begin{document}

\title{Construction and description of the stationary measure of weakly dissipative dynamical systems}

\author{Itzhak Fouxon}
\affiliation{Raymond and Beverly Sackler School of Physics and Astronomy,
Tel-Aviv University, Tel-Aviv 69978, Israel}
\begin{abstract}

We consider the stationary measure of the dissipative dynamical system in a finite volume. A finite dissipation, however small, generally makes the measure singular, while at zero dissipation the measure is constant. Thus dissipative part of the dynamics is a singular perturbation producing an infinite change in the measure. 
This is a result of the infinite time of evolution that enhances the small effects of dissipation to form singularities. We show how to deal with the singularity of the perturbation and describe the statistics of the measure. We derive all the correlation functions and the statistics of "mass" contained in a small ball. The spectrum of dimensions of the attractor is obtained.
The fractal dimension is equal to the space dimension, while the information dimension is equal to the Kaplan-Yorke dimension.

\end{abstract}
\pacs{05.45.Df} \maketitle

\section{Introduction}

The dynamical system is one of the most studied paradigms in science that arises in many different contexts and has roots in the classical mechanics. Within that paradigm, the state of the system at time $t$ is represented by a vector $\bm x(t)$ in a $d-$dimensional "phase space". It is assumed that the evolution of that vector is determined by its position via a smooth velocity field $\bm V[t, \bm x]$,
\begin{eqnarray}&&
\frac{d\bm x}{dt}=\bm V\left[t, \bm x(t)\right].  \label{basic1}
\end{eqnarray}
The above equation can be considered as the definition of the dynamical system. For a generic $\bm V$ the trajectories of the system are chaotic and
one shifts the analysis from the study of a single trajectory to the study of the phase space density. The latter obeys the continuity (Liouville) equation which time-independent solutions may describe a steady state of the system. For incompressible velocity fields, with $\nabla\cdot\bm V=0$, a constant solves the continuity equation. If the system is mixing \cite{Ma,Dorfman}, then the constant is the solution that describes the steady state of the system. Here and below we assume the flow occurs in a finite volume so the constant distribution is normalizable. The steady state described by a constant phase space density is, in particular, the case of the classical equilibrium statistical mechanics. There the uniform phase space density - the microcanonical distribution - describes the steady state of a closed system.

In many effective descriptions of the dynamics of the open systems, the dynamical system is dissipative. Dissipation leads to the non-conservation of the Gibbs entropy which implies breaking of the Liouville theorem on the conservation of the phase-space volumes by the flow. The non-conservation of the volumes signifies a finite divergence of the velocity field, so that a dissipative dynamical system can be defined as the system (\ref{basic1}) with a non-vanishing divergence of $\bm V$. One can still consider the time-independent solutions to the continuity equation to
describe the steady state of the open system that exchanges entropy with the environment. A constant is however no longer a solution to the continuity equation, because of the non-vanishing divergence of velocity. Finding the time-independent solutions to the continuity equation becomes a non-trivial problem and generally it is not possible to describe the stationary measure for a given flow $\bm V[t, \bm x]$ \cite{Dorfman,Ruelle1}. For a generic $\bm V$ there are no smooth time-independent solutions. The absence of smooth solutions to the coninuity equation is generally necessary to model open systems that exchange entropy with the environment. Indeed, note that the Gibbs entropy is a functional of the phase space density. If the latter would be smooth in the steady state, that would imply that the Gibbs entropy of the system is constant in that state. This would contradict the entropy exchange with the environment \cite{Ruelle1}. The absence of a smooth solution does not present a problem for the description in terms of the phase space density. The latter need not be smooth, since the measurable quantities are the integrals of that density with smooth functions, and not the density itself. Furthermore, the original form of the mass conservation law is integral, while the continuity equation follows from it, assuming the density is differentiable.

The time-independent solutions of the continuity equation that satisfy the equation in the sense of integrals are called "weak solutions". They have singularities of the $\delta-$function type so the integrals with smooth functions are well-defined. For these solutions the conclusion on the conservation of the Gibbs entropy for the time-independent solution of the continuity equation does not work: the Gibbs entropy is not defined for a singular phase space density. Nevertheless, in many cases the time derivative of the Gibbs entropy is well-defined for the singular density solution and is given by a finite constant. This is the property one expects from non-equilibrium steady states.

A well-established situation where singular solutions to the continuity equation allow quite detailed description is the Sinai-Ruelle-Bowen (SRB) measure \cite{Ruelle1,Dorfman,Sinai,Bowen}. These exist under certain assumptions on the velocity field $\bm V(\bm x)$.
This measure is supported on a zero-volume, multi-fractal set in space - the "strange attractor". The support has the property that it is approached by the system trajectories asymptotically at large times. Due to the zero volume
of the support the measure is singular. The measure reflects the long-time behavior of the dynamical trajectories and it allows to calculate the time averages of the dynamical variables given by functions on the phase space.

In this work we find explicitly the stationary measure (density) that describes the steady state of weakly dissipative systems. This is the
case of weakly compressible $\bm V[t, \bm x]$, where the potential component is much smaller than the solenoidal one.
We construct the representation of the measure with the help of $\bm V$ and provide the validity conditions of the representation. We describe completely the statistics of that measure as defined by the spatial averaging. Our results pertain both to a time-independent deterministic flow and to a time-dependent flow, which statistics defined by the spatial averaging is stationary.

Weak compressibility means that the Liouville theorem "almost" holds, so the Gibbs entropy is almost conserved and the system is near equilibrium.
The stationary measure is expected to be close to a constant. Nevertheless, at however small dissipation, there are always dynamical variables for which the average is not close to the equilibrium average obtained from the microcanonical ensemble. The large evolution time compensates the smallness of the dissipation and the singular steady state density is significantly different from the smooth constant density.

To see how the limit of small compressibility fits the dynamics, consider the decomposition of the velocity field $\bm V$ into the solenoidal and the potential parts
\begin{eqnarray}&&
\bm V\left[t, \bm x\right]=\bm u\left[t, \bm x\right]+\epsilon \bm v\left[t, \bm x\right], \ \ \nabla\cdot \bm u=0,\label{basic12}
\end{eqnarray}
where $\epsilon$ is the small parameter of our analysis. At $\epsilon=0$ for a mixing system, the evolution of a small volume results in a volume
which coarse-graining over the infinitesimally small scale fills the whole phase space. We will see that at $\epsilon>0$ the scale $l(\epsilon)$
over which one should make a coarse-graining to cover the whole space becomes a finite scale vanishing with $\epsilon$.


\section{Kaplan-Yorke dimension of the attractor}

We introduce the mapping ${\bf q}(t, {\bf x})$ that gives the position of the system trajectory at time $t$ provided it was initially at $\bm x$. This satisfies
\begin{eqnarray}&&
\partial_t {\bf q}(t, {\bf x})={\bf V}[t, {\bf
q}(t, {\bf x})],\ \ {\bf q}(0, {\bf x})={\bf x}.
\label{main}
\end{eqnarray}
The flow is assumed to be confined to a finite volume, which we set equal to unity with no loss. Correspondingly either the velocity field has zero normal component at the boundary or the periodic boundary conditions are assumed. We study both the case of a time-independent deterministic flow $\bm v(t, \bm x)=\bm v(\bm x)$ and the case of a time-dependent flow $\bm v(t, \bm x)$ which is stationary with respect to the statistics defined by the spatial averaging.

Much insight into the behavior of the system trajectories is provided by the sum of the Lyapunov exponents \cite{Oseledets}. The sum of the forward in time Lyapunov exponents $\sum\lambda_i^+(\bm x)$ and the backward in time Lyapunov exponents $\sum\lambda_i^-(\bm x)$ describe the evolution of infinitesimal volumes in the phase space both forward and backward in time,
\begin{eqnarray}&&
\sum\lambda_i^+(\bm x)\equiv \lim_{t\to\infty}\frac{1}{t}\ln\det\frac{\partial \bm q(t, \bm x)}{\partial \bm x},\label{a12}\\&&
\sum\lambda_i^-(\bm x)\equiv \lim_{t\to \infty}\frac{1}{t}\ln\det\frac{\partial \bm q(-t, \bm x)}{\partial \bm x}.\label{a122}
\end{eqnarray}
The Jacobian can be written explicitly as
\begin{eqnarray}&&\!\!\!\!\!\!\!\!\!\!\!\!\!\!
\det\frac{\partial \bm q(t, \bm x)}{\partial \bm x}=\exp\left[\int_0^t W[t, \bm q(t, \bm x)]\right],\ \ W\equiv \nabla\cdot \bm V \label{a4}.
\end{eqnarray}
The above formulas show the sums of the Lyapunov exponents $\sum\lambda_i^{\pm}$ can be written as a time-average of a function on the phase space
\begin{eqnarray}&&
\sum\lambda_i^+(\bm x)= \lim_{t\to\infty}\frac{1}{t}\int_0^t W\left[t', \bm q(t', \bm x)\right]dt'.\label{a1222}\\&&
\sum\lambda_i^-(\bm x)=-\lim_{t\to\infty}\frac{1}{t}\int_{-t}^0 W\left[t', \bm q(t', \bm x)\right]dt'. \label{a121212}
\end{eqnarray}
A similar expression holds for $\sum\lambda_i^-$. This special form was used  in \cite{FF,IF} to derive a Green-Kubo type formula,
\begin{eqnarray}&&
-\int \sum\lambda_i^{\pm}(\bm x)d\bm x=
\pm \int_0^{\pm\infty}dt \langle W(0)W(t)\rangle_L,\nonumber\\&&
\langle W(0)W(t)\rangle_L=\int W(0, \bm x)W[t, \bm q(t, \bm x)] d\bm x. \label{a1}
\end{eqnarray}
Note that no similar writing is possible for other combinations of $\lambda_i^{\pm}$ and the above representation is unique for $\sum\lambda_i^{\pm}$. When the integral converges, the formula holds for an arbitrary smooth $\bm V[t, \bm x]$, in particular, far from equilibrium at an arbitrary compressibility of $\bm V$, or $\epsilon\gtrsim 1$. The subscript $L$ stresses the correlation function is defined with the help of the Lebesgue measure rather than the stationary measure of the system.

We now study the $\epsilon\to 0$ limit of Eqs.~(\ref{a1}). We introduce $w(t, \bm x)\equiv \nabla\cdot \bm v$, so $W(t, \bm x)=\epsilon w(t, \bm x)$. At weak compressibility, to leading order, one may substitute $\bm q(t, \bm x)$ in Eq.~(\ref{a1}) by the trajectories of the $\epsilon=0$ flow $\bm X(t, \bm x)$,
\begin{eqnarray}&&
\partial_t {\bm X}(t, \bm x)=\bm u[t, \bm X(t, \bm x)],\ \ \bm X(0, \bm x)=\bm x.
\end{eqnarray}
We define the "microcanonical" correlation function by
\begin{eqnarray}&&
\langle w(0)w(t)\rangle\equiv \int w(0, \bm x)w[t, \bm X(t, \bm x)] d\bm x.
\end{eqnarray}
By incompressibility $\langle w(0)w(t)\rangle=\langle w(0)w(-t)\rangle$ so that
\begin{eqnarray}&&\!\!\!\!\!\!\!
-\int \sum\lambda_i^{\pm}(\bm x)d\bm x=\frac{\epsilon^2}{2}\int_{-\infty}^\infty \langle w(0)w(t)\rangle dt=\frac{\epsilon^2 E(0)}{2},\nonumber
\end{eqnarray}
where $E(0)$ is the spectrum of $\omega[t, \bm X(t, \bm r)]$ at zero frequency, $E(0)\geq 0$. We observe that at small $\epsilon$ the sums are equal and behave as $\epsilon^2$, though $W$ behaves as $\epsilon$. To see how the extra $\epsilon$ factor arises, note that the sums of the Lyapunov exponents could be obtained as the spatial average of the divergence of velocity on the system trajectory, $\epsilon \langle w\left[t, \bm q(t, \bm x)\right] \rangle$, since the divergence equals the local logarithmic growth rate of infinitesimal volumes. However, $\epsilon \langle w\left[t, \bm X(t, \bm x)\right]\rangle$, that could be the leading order approximation, vanishes identically by
\begin{eqnarray}&&
\langle w\left[t, \bm X(t, \bm x)\right]\rangle=\int w\left[t, \bm X(t, \bm x)\right]d\bm x\nonumber\\&&
=\int w\left[t, \bm x\right]d\bm x=0.
\end{eqnarray}
Above we noted that the spatial integral of the divergence of $\bm v$, which is equal to the surface integral of $\bm v$, vanishes either
due to the periodic boundary conditions or due to the vanishing velocity at the boundary, since the vanishing must hold separately for the
potential and solenoidal components of the velocity.

The sum of the Lyapunov exponents determines the entropy production rate in the non-equilibrium steady state \cite{Ruelle1,FF} and the
above result appears quite important from the general viewpoint. The proof that $\int\sum\lambda_i^{\pm}\leq 0$ is tantamount to the
statement of the second law of thermodynamics for the considered framework \cite{Ruelle1,FF}.

Assuming that the system is generic and there is no degeneracy we have $E(0)>0$. Then, necessarily, the above expressions for space-averaged sums of the Lyapunov exponents imply that, at least in some regions of the phase space, the sums of the Lyapunov exponents are negative. This has important implications for the evolution of the smooth phase space density $n$ that
obeys the mass conservation relation 
\begin{eqnarray}&&
n\left[t, \bm q(t, \bm x)\right]\det\frac{\partial \bm q(t, \bm x)}{\partial \bm x}=n(0, \bm x). \label{mass}
\end{eqnarray}
It follows from the above relation and Eqs.~(\ref{a12})-(\ref{a122}) that sums of the Lyapunov exponents also determine the evolution of the density
\begin{eqnarray}&&
\lim_{t\to\infty}\frac{1}{t}\ln\frac{n[t, \bm q(t, \bm x)]}{n(0, \bm x)}=-\sum\lambda_i^+(\bm x),\nonumber\\&&
\lim_{T\to\infty}\frac{1}{T}\ln\frac{n(0, \bm x)}{n[-T, \bm q(-T, \bm x)]}=\sum\lambda_i^-(\bm x),
.\label{a2}
\end{eqnarray}
It follows from the second equation that if we fix an initial condition for the density in the remote past at $t=-T$ and take the limit of the infinite evolution time $T\to\infty$, then in the region where $\sum\lambda_i^-(\bm x)<0$ the density decays exponentially to zero producing a void with a finite volume. Complementarily, the density on the trajectories that issue from the region with $\sum\lambda_i^+(\bm x)<0$ grows exponentially, producing infinite density in the infinite evolution time. Thus for non-degenerate systems with $E(0)>0$ the stationary density (assuming it can be obtained as a result of infinite time of evolution, see below) will have both infinities and zeros, i. e. it is singular.

The representation (\ref{a1222}) and a similar representation for $\sum\lambda_i^-$ suggest that in many cases the limits
$\sum\lambda_i^{\pm}(\bm x)$ do not depend on $\bm x$ and are the same for almost all $\bm x$, so that $\sum\lambda_i^{\pm}(\bm x)=\Theta<0$
possibly except for a set with zero
measure. While we leave the determination of the conditions when this holds true for further work, here we mention that
if this is the case, then the above consideration shows that the stationary density is zero almost everywhere and the volume of its support is zero.
Thus if $\sum\lambda_i^+(\bm x)=\Theta$ is a constant for almost all $\bm x$, then the trajectories approach at large times a set with zero volume - "a strange attractor". For a time-dependent $\bm V(t,\bm x)$ this attractor varies in time and it could be more proper to call it "a random attractor".

We now find the Kaplan-Yorke codimension $C_{KY}$ of the attractor \cite{KY}.
We notice that to the leading order in $\epsilon$ the exponents $\lambda_i^+$ are equal to the exponents $\lambda_i$ of the $\epsilon=0$ incompressible flow $\bm u$. Therefore $\sum_{i=1}^{d-1}\lambda_i^+\approx -\lambda_d>0$ while $\sum\lambda_i^+<0$. Here we use that by incompressibility $\sum\lambda_i=0$ and assume $\lambda_i$ do not vanish identically (this implies $\lambda_d<0$ by $\sum \lambda_i=0$). By definition of $C_{KY}$, it follows from $\sum_{i=1}^{d}\lambda_i^+<0$ that
\begin{eqnarray}&&
C_{KY}=\frac{\sum \lambda_i^+}{\lambda_d^+}\approx \frac{\epsilon^2 E(0)}{2|\lambda_d|}.
\end{eqnarray}
The above formula describes the Kaplan-Yorke codimension of the attractor both for time-independent $\bm V\left[\bm x\right]$ and for time-dependent $\bm V\left[t, \bm x\right]$. In the latter case the attractor evolves in time preserving its fractal dimensions, see below. Thus a lot of information on the resultant large-time behavior of Eqs.~(\ref{basic1})-(\ref{basic12}) is available thanks to the Green-Kubo type formulas (\ref{a1}).

The Kaplan-Yorke codimension vanishes as $\epsilon^2$ at $\epsilon\to 0$ so that the attractor is almost space-filling in the sense of the Kaplan-Yorke dimension. However, the attractor has a whole infinity of fractal dimensions which study reveals further details on the structure
of the attractor. These can be established by studying the detailed structure of the stationary measure of the system.

\section{A general representation of the stationary measure}

We construct an explicit representation of the stationary measure in terms of the velocity $\bm V$. The analysis in this chapter is
general and, not using the assumption of weak dissipation, it applies to any dynamical system.
The stationary measure $n_s$ is a time-independent solution to the continuity equation
\begin{eqnarray}&&
\partial_t n+\nabla\cdot(n\bm V)=0,\ \ \label{continuity}
\end{eqnarray}
where we first consider time-independent $\bm V(\bm x)$ and then generalize to time-dependent fields with stationary statistics. Generally, for
dissipative systems one has to consider weak solutions to the above equation as no smooth solution is possible. To consider the possibility of
smooth time-independent solutions, we assume a smooth solution exists
and show the conditions on $\bm V$ that this assumption implies. A smooth solution would have to be invariant under the time evolution
(\ref{mass}), i. e. it would have to obey
\begin{eqnarray}&&
n_s\left[\bm q(t, \bm x)\right]\det\frac{\partial \bm q(t, \bm x)}{\partial \bm x}=n_s(\bm x). \label{mass1}
\end{eqnarray}
Taking the logarithm of both sides divided by $n_s\left[\bm q(t, \bm x)\right]$ and applying the limit $t\to\infty$
we obtain
\begin{eqnarray}&&
\sum \lambda_i^+(\bm x)=\lim_{t\to\infty}\frac{1}{t}\ln\left(\frac{n_s(\bm x)}{n_s\left[\bm q(t, \bm x)\right]}\right)=0, \label{mass2}
\end{eqnarray}
where the last equality follows from the fact that the ratio $n_s\left[\bm q(t, \bm x)\right]/n_s(\bm x)$ has both a finite maximum and a finite positive minimum, by the assumption of the smoothness of $n_s(\bm x)$ and the finiteness of the volume of the phase space, so that the limit is the one of a bounded quantity over time. Thus if the system admits smooth time-independent solutions to the continuity equation, then it must have
$\sum \lambda_i^+(\bm x)\equiv 0$ at every $\bm x$. In particular, $\int \sum \lambda_i^+(\bm x)d\bm x=0$, which corresponds to $E(0)=0$, cf. \cite{IF}. We conclude that smooth time-independent solutions to the continuity equation may exist only if $E(0)=0$.

Thus, confining ourselves to the study of the non-degenerate situation $E(0)>0$,  we must search for non-smooth time-independent solutions
to the continuity equation, i. e. distributions that provide a weak solution $n_s$ to the equation. A weak solution of the continuity equation satisfies the equation in the integral sense. This solution is a distribution of mass $m_l(\bm x)$ that gives the mass contained in the ball
with asymptotically small radius $l$ centered at $\bm x$. This mass distribution defines a generalized density by
\begin{eqnarray}&&
n_s(\bm x)=\lim_{l\to\infty}\frac{m_l(\bm x)}{V_l},
\end{eqnarray}
where $V_l$ is the volume of the ball with radius $l$ in $d$ dimensions. The above limit in the considered case where the solution is supported
on the multi-fractal attractor is either zero or infinity and thus is not well-defined. However the limit is well-defined in the sense of
distributions. The phase space average
\begin{eqnarray}&&
\langle f\rangle \equiv \lim_{l\to\infty}\frac{1}{V_l}\int f(\bm x)m_l(\bm x)d\bm x,
\end{eqnarray}
is well-defined and it is conserved by the dynamics if the mass distribution corresponds to the time-independent solution to the continuity
equation when the average is time-independent $\langle f\rangle=\langle f(t)\rangle$, or
\begin{eqnarray}&&
\!\!\!\!\!\!\!\lim_{l\to\infty}\frac{1}{V_l}\int f(\bm x)m_l(\bm x)d\bm x\!=\!\!\lim_{l\to\infty}\frac{1}{V_l}\int f\left[\bm q(t, \bm x)\right]m_l(\bm x)d\bm x.\nonumber
\end{eqnarray}
The distribution if called a weak solution if the averages are conserved for any smooth $f$.
If $n_s(\bm x)$ were smooth, the above definition would reduce to $\nabla\cdot\left[n_s\bm V\right]=0$, as can be seen differentiating $\int f\left[\bm q(t, \bm x)\right]n_s(\bm x)d\bm x$ with respect to time, putting $t=0$ and integrating by parts.
Below we will denote the solution by the density $n_s(\bm x)$ implying the corresponding distribution of the mass and
write symbolically $\langle f\rangle=\int f(\bm x)n_s(\bm x)d\bm x$.
We conclude that a weak solution is a physically meaningful quantity allowing to calculate averages of dynamical variables.

One expects that quite generally $n_s$ can be obtained as the distribution that results from an arbitrary smooth initial condition in the limit of the infinite time of evolution. We set with no loss $n(t=-T)=1$ and consider the limit $T\to\infty$. Since the continuity equation is the differential form of the mass conservation law (\ref{mass}), then it follows from the latter equation and Eq.~(\ref{a4}) that the solution for $n(t=-T)=1$ is
\begin{eqnarray}&&
n(t=0, \bm x)=\exp\left[-\int_{-T}^0 W[\bm q(t, \bm x)]dt\right].
\end{eqnarray}
where we specified to the time-independent velocity field.
We now establish the conditions under which the above expression produces in the limit $T\to\infty$ a weak solution $n_s$ to the continuity
equation. We consider the derivative of $\langle f(t)\rangle=\int f(\bm x) n(t, \bm x) d\bm x$ at $t=0$ for an arbitrary smooth function $f$. The derivative $\langle {\dot f}(0)\rangle=-\int d\bm x f \nabla\cdot[n(t=0)\bm V]$  can be written as
\begin{eqnarray}&&
\!\!\!\!\!\langle {\dot f}(0)\rangle=-\!\int f d\bm x\nabla\cdot \left[\bm V(\bm x)\exp\left[-\int_{-T}^0 W[\bm q(t, \bm x)dt\right]\right]\nonumber\\&&\!\!\!\!\!
=\int d\bm x\exp\left[-\int_{-T}^0 W[t, \bm q(t, \bm x)dt\right]\bm V(\bm x)\cdot\nabla f(\bm x)\nonumber.
\end{eqnarray}
where we assumed the vanishing of the boundary terms in integration by parts, as realized either by vanishing $\bm v$ on the boundaries or by the periodic boundary conditions. We notice that $d\bm x\exp\left[-\int_{-T}^0 W[t, \bm q(t, \bm x)dt\right]=d\bm y$ where $\bm y=\bm q(-T, \bm x)$ so the inverse change of variables $\bm x=\bm q(T, \bm y)$ transforms the integral into
\begin{eqnarray}&&
\!\!\!\!\!\!\!\!\int d\bm y\left[\bm V\cdot\nabla f(\bm x)\right]|_{\bm x=\bm q(T, \bm y)}
=\int d\bm y \bm V(\bm y)\cdot\nabla_{\bm y} f[\bm q(T, \bm y)],\nonumber
\end{eqnarray}
where we noticed the formula
\begin{eqnarray}&&
V_i[\bm q(T, \bm y)]=V_j(\bm y)\frac{\partial q_i(t, \bm y)}{\partial y_j}.  \label{vel}
\end{eqnarray}
The above equation expresses the fact that if we consider two trajectories issuing from the point $\bm y$ and the point $\bm y+\bm v(\bm y)\Delta t$, shifted along the trajectory $\bm q(t, \bm y)$, where $\Delta t$ is infinitesimal, then these trajectories at time $t$ must be at $\bm q(t, \bm y)$ and $\bm q(t+\Delta t, \bm y)$. Thus the distance between the trajectories at time $t$ is on the one hand equal to $\Delta t\partial_t q_i(t, \bm y)=\delta t v_i[\bm q(t, \bm y)]$ and, on the other hand, it is determined by the derivative of $\bm q(t, \bm y)$ as initial distance $v_j(\bm y)\Delta t$ times the derivative $\partial_j q_i(T, \bm y)$, summed over $j$. The comparison of the two expressions gives Eq.~(\ref{vel}).

Finally, integrating by parts and again assuming the vanishing contribution of the boundaries, we find
\begin{eqnarray}&&
\!\!\!\!\!\!\!\!\!\!\langle {\dot f}(0)\rangle
=-\int W(\bm y)f[\bm q(T, \bm y)]d\bm y=-\langle W(0)f(T)\rangle.\nonumber
\end{eqnarray}
The above relation is exact and it holds independently of the properties of $\bm V(\bm x)$. We now make the assumption that the correlations decay
so that
\begin{eqnarray}&&
\lim_{t\to\infty}\langle W(0)f(T)\rangle=\langle W(0)\rangle\langle f(T)\rangle=0,
\end{eqnarray}
where we used that $\int W(\bm y) d\bm y=0$ both for periodic boundary conditions, or for velocity vanishing on the boundaries.
This assumption guarantees $\langle {\dot f}(0)\rangle=0$ in the limit $T\to\infty$.
We conclude that the distribution
\begin{eqnarray}&&
n_s=\lim_{T\to\infty}\exp\left[-\int_{-T}^0 W[t, \bm q(t, \bm x)]dt\right],\label{solution}
\end{eqnarray}
where the limit is taken after taking the integral, provides a weak stationary solution to the continuity equation provided the correlations decay in time and $\langle W(0)f(T)\rangle$ tends to zero at large $T$ for any smooth function.
Below we assume that the correlation decay condition is satisfied and Eq.~(\ref{solution}) gives the correct stationary measure of the system.

The above derivation can be generalized to the case of the time-dependent velocity field with stationary statistics, following exactly the same steps that were provided for a similar generalization in \cite{FF}. One finds that that the same condition of decay of correlations guarantees the
(\ref{solution}) is the weak stationary solution to the continuity equation.

The solution (\ref{solution}) provides the measure appearing in
the SRB theorem \cite{Dorfman} on the equality of time and phase-space averages for dynamical variables $f$,
\begin{eqnarray}&&
\!\!\!\!\!\!\!\!\!\!\lim_{T\to\infty} (1/T)\int_0^T f[\bm q(t, \bm x)]dt=\int f(\bm r) n_{s}(\bm r)d\bm r,
\end{eqnarray}
provided the time-average is independent of $\bm x$ for almost every $\rm x$. The proof is based on the identity
$\lim_{T\to\infty} \int f[\bm q(T, \bm x)]d\bm x=\int
f(\bm r) n_{s}(\bm r)d\bm r$.

We stress again that the representation derived here does not use the assumption of weak dissipation and can be used for general studies of the
stationary measures of dynamical systems. Below we provide several general relations on $n_s$.

\section{Build-up of singular density, a sum rule and pair correlation function}

Here we analyze how the singularities in $n_s$ build up in the course of the evolution. We also derive a sum rule connecting the trajectories
$\bm q(t, \bm x)$ and the velocity $\bm V$. We provide an expression for the correlation function of $n_s$. The main assumption of the analysis below is that $W[t, \bm q(t, \bm x)]$ has a finite correlation time $\tau_c$. This assumption implies the assumption of decay of correlations introduced in the previous section.

We observe that Eq.~(\ref{solution}) gives $n_s$ as the exponent of a sum of a large, asymptotically infinite, number $\sim T/\tau_c$ of uncorrelated random variables. This implies the limit $T\to\infty$ in $n_{st}$ generally brings a pointwise singular measure in agreement with the trajectories' accumulation on the strange attractor. The build up of the singularities is characterized by considering the change in the moments of the single-point density $n(t=0)$ as $T$ grows
\begin{eqnarray}&&
\langle n^k(t=0)\rangle=\left\langle\int\exp\left[-k\int_{-T}^0 W[t, \bm q(t, \bm x)]dt\right]\right\rangle.\nonumber
\end{eqnarray}
By the cumulant expansion theorem (see e. g. \cite{Ma}), the logarithms $I_k\equiv \ln \langle n^k(t=0)\rangle$ can be written as
\begin{eqnarray}&&
\!\!\!\!\!\!\!\!I_k(T)=\sum_{m=1}^{\infty}\frac{(-k)^m}{M!}\left\langle\left(
\int_{-T}^0 W[t, \bm q(t, \bm x)]dt\right)^m\right\rangle_c.\label{a2a2}
\end{eqnarray}
By the assumption of the finite correlation time $\tau_c$ of $W[t, \bm q(t, \bm x)]$, all the cumulants are proportional to $T$
at large $T$, so the functions
\begin{eqnarray}&&
\gamma(k)\equiv \lim_{T\to\infty}\frac{I_k(T)}{T},
\end{eqnarray}
are well-defined, see \cite{BFF}. By the Holder inequality the function $\gamma(k)$ is convex. It also obeys $\gamma(0)=0$ and $\gamma(1)=0$ due to the
conservation of the total mass $M$, so $\langle n\rangle=const=M$. Furthermore $\gamma'(0)=\sum \lambda_i^-<0$ (we assume $E(0)>0$) as can be verified by direct differentiation, using Eq.~(\ref{a121212}). Convexity then implies that $\gamma(k)\leq 0$ for $0<k<1$ and $\gamma(k)\geq 0$ otherwise. Thus in the non-degenerate case where sharp inequalities hold we have that the moments of $n_s$ are either plus infinity or zero characterizing the singularity of the density. For example the root mean square deviation $\langle n_s^2\rangle/\langle n_s\rangle^2-1$ is infinite.

The mass conservation  $I_1=0$ implies by Eq.~(\ref{a2a2}) that
\begin{eqnarray}&&
\sum_{m=1}^{\infty}\frac{(-1)^m}{M!}\left\langle\left(
\int_{-T}^0 W[t, \bm q(t, \bm x)]dt\right)^m\right\rangle_c=0.\label{a2a2a2a2}
\end{eqnarray}
The above sum rule holds for any $T$ and it
means that the relation between the velocity divergence and the trajectories is not arbitrary, but is should be such as
to conserve the overall volume of the phase-space.

The sum rule seems to be new. Its main use is that it will allow us to write down finite expressions for the correlation functions of $n_s$.
We introduce a short-hand notation $\rho(-T, \bm x)=-\int_{-T}^0 W[t, \bm q(t, \bm x)]dt$ so that the sum rule reads
\begin{eqnarray}&&
\sum_{n=1}^{\infty}\frac{1}{n!} \left\langle \rho^n
\right\rangle_c=0.
\label{s1}\end{eqnarray}
We now consider the pair correlation function of $n_s$. While the single-point moments of the density do not have a finite (non-zero) limit
at $T\to\infty$, the correlation function
\begin{eqnarray}&&
\!\!\!\!\!\!\!\!\!\langle n_{s}(0)n_{s}(\bm x)\rangle\!\!=\!\exp\left[g(\bm x)\right],\  \  g(\bm x)\!\equiv\!\!\lim_{T\to\infty} \ln \langle\exp[\rho_1+\rho_2]\rangle.\nonumber
\end{eqnarray}
has a finite limit at $T\to\infty$. Here we defined $\rho_1=-\int_{-T}^0 w[t, \bm q(t, \bm x)]dt$ and $\rho_2=-\int_{-T}^0 w[t, \bm q(t, 0)]dt$. We will suppress
the arguments in $\rho_i$.
To derive the limit we apply the cumulant expansion theorem to the above equation which gives
\begin{eqnarray}&&
g(\bm x)=\lim_{T\to\infty}\sum_{n=1}^{\infty}\frac{1}{n!}\langle[\rho_1\!+\!\rho_2]^n\rangle_c
,\label{general}
\end{eqnarray}
Though it might seem the series above diverges, it does not due to the sum rule (\ref{s1}) that implies that
only mixed terms contribute,
\begin{eqnarray}&&
\!\!\!\!\!\!\!\!\!\!g(\bm x)=\lim_{T\to\infty}\sum_{n=2}^{\infty}
\frac{\langle[\rho_1\!+\!\rho_2]^n\rangle_c\!-\!\langle \rho_1^n\rangle_c\!-\!\langle \rho_2^n\rangle_c}{n!}
,\label{a2222}
\end{eqnarray}
In the mixed terms the correlations decay due to the separation of the trajectories. Consider the $n=2$ term $Q$ that can be written
as
\begin{eqnarray}&&
\!\!\!\!\!\!\!\!\!\!\!\!\!\!\!\!\!\!
Q\!\equiv\! \lim_{T\to\infty}\int_{-T}^0 dt_1dt_2\langle
W[t_1, \bm q(t_1, 0)]W[t_2, \bm q(t_2, \bm x)]\rangle_c. \label{quadr}
\end{eqnarray}
For $x=0$ the trajectories $\bm q$ stick together forever and the integral diverges in correspondence with the divergence of $\langle n_{s}^2\rangle$. In contrast, at $\bm x\neq 0$, the trajectories diverge backward in time making the integral convergent: for $t_1\to-\infty$ the correlation function in the integral is negligible. Since the divergence
is exponential, then the time $t_*$ during which $|\bm q(t, \bm x)-\bm q(t, 0)|$ is much smaller than the scale of variations
of $w(t, \bm x)$, so that $w[t, \bm q(t, \bm x)]\approx w[t, \bm q(t, \bm x)]$ diverges logarithmically at small $r$, i. e. $t_*\propto |\ln (r)|$.
Similar divergence holds for $n>2$ terms in the expression for $\ln\langle n_{s}(0)n_{s}(\bm x)\rangle$. In this way one finds that
$\langle n_{s}(0)n_{s}(\bm x)\rangle$ diverges as a power-law at small $x$. Here we will be interested in the case of small compressibility or weak dissipation.

\section{The statistics of $n_s$ at weak dissipation}

We show that in the limit of small $\epsilon$ a complete description of the properties of $n_s$ given by Eq.~(\ref{solution}) is possible. The analysis is written for $\bm V=\bm V(\bm x)$ while generalizations to $\bm V=\bm V(t, \bm x)$ case are either straightforward or follow \cite{FF,IF} literally.

\subsection{Calculation of the correlation functions of $n_s$}

In the limit of small compressibility $\epsilon\ll 1$ the series (\ref{a2222}) is dominated by $n=2$ term. Furthermore, to the leading
order one can substitute $\bm q(t, \bm x)$ in Eq.~(\ref{quadr}) by $\bm X(t, \bm x)$. We find
\begin{eqnarray}&&
\!\!\!\!\!\!\!\!\!\!
g(\bm x)=\epsilon^2 \int_{-\infty}^0\!\!\!\!\! dt_1dt_2\langle
\omega[{\bf X}(t_1, 0)] \omega[{\bf X}(t_2, {\bf x})]\rangle+o(\epsilon),\!\!\!\!\!\!
\label{structure} 
\\&&\!\!\!\!\!\!\!\!\!\!
\langle n_s(0)n_s(\bm x)\rangle\!\!\approx\! \exp\left[\epsilon^2\!\! \int_{-\infty}^0\!\!\!\!\! dt_1dt_2\langle
\omega[{\bf X}(t_1, 0)] \omega[{\bf X}(t_2, {\bf x})]\rangle
\right].\nonumber \end{eqnarray}
Above we omitted the subscript c since for incompressible
flow $\langle w[{\bf X}(t, {\bf x})]\rangle=\langle w({\bf x})\rangle=0$.
The condition of applicability of Eqs.~(\ref{structure}) is the negligibility of $n>2$ terms in Eq.~(\ref{general}).

An important result is obtained by repeating the procedure used for $\langle n_s(0)n_s(\bm x)\rangle$ for higher order correlation
functions. Applying the cumulant expansion to
\begin{eqnarray}&&\!\!\!\!\!\!\!\!\!\!\!
\ln \langle n(\bm x_1)n(\bm x_2)..n(\bm x_N)\rangle\!\!=\!\!\lim_{T\to \infty}\!\!\ln\! \left\langle \!\exp\left[\sum_{i=1}^N\rho(-T, \bm x_i)\right]\right\rangle,\nonumber
\end{eqnarray}
one finds
\begin{eqnarray}&&
\!\!\!\!\!\!\!\!\!\!\!\!\!\!\!\!\!\!\!\!\ln \langle n_{s}(\bm x_1)n_{s}(\bm x_2)..n_{s}(\bm x_N)\rangle
=\sum_{i>j}g({\bf x}_{i}-{\bf
x}_j). \label{bueno}\end{eqnarray}
We conclude that $n_{s}$ has a log-normal statistics which is completely determined by a single structure function
$g(\bm x)$.

The structure function has a universal behavior at small $x$. It diverges at $x=0$  
because $\bm X(t, 0)$ and $\bm X(t, \bm x)$ do not separate and spend infinite time together:
$\int dt_1dt_2\langle\omega[{\bf X}(t_1, 0)] \omega[{\bf X}(t_2, 0)]\rangle=\infty$. This
corresponds to $\langle n_s^2\rangle=\infty$ mentioned earlier.
At a small but finite $x$ the integral in $g(\bm x)$ becomes finite. The distance $R(t_1)\equiv |{\bf X}(t_1, \bm x)-{\bf X}(t_1, 0)|=R(t_1)$ grows
exponentially with $|t_1|$ with the rate equal to the principal Lyapunov exponent $|\lambda_d|$ of the time-reversed flow $\bm u$. As a result,
when $R(t_1)$ becomes comparable with the spatial correlation length $\eta$ of $w$, the integrand in $g(\bm x)$ starts to decay and
the integral in $g(\bm x)$ converges. However, the trajectories stay together during a time that diverges logarithmically at $x\to 0$, so
that the divergence of $g(\bm x)$ at small $x$ is logarithmic. Here we give a somewhat heuristic derivation of the divergent part of $g(\bm x)$,
postponing the more precise derivation to the following subsections.

To single out the divergent part of $g(\bm x)$ we first change
the interval of integration over $t_2$:
\begin{eqnarray}&&\!\!\!\!\!\!\!\!\!\!\!\!\!\!\!\!\!\!
g(\bm x)\approx \epsilon^2 \int_{-\infty}^0\!\!\!\!\! dt_1\int_{-\infty}^{\infty} dt_2\langle
\omega[{\bf X}(t_1, 0)] \omega[{\bf X}(t_2, {\bf x})]\rangle.  \end{eqnarray}
The above approximation neglects the finite contribution of $0<t_2\lesssim \tau_c$ versus the contribution becoming infinite
at $x\to 0$. Next, we introduce $F(t_1)$ by $g(\bm x)\approx \epsilon^2 \int_{-\infty}^0 F(t_1) dt_1$, so
\begin{eqnarray}&&\!\!\!\!\!\!\!\!\!\!\!\!\!\!\!\!\!\!
F(t, \bm x)=\int_{-\infty}^{\infty} dt'\langle
\omega[\bm X(t, \bm x)] \omega[{\bf X}(t', 0)]\rangle.  \end{eqnarray}
The integral converges over the characteristic times $|t_2-t_1|\lesssim \tau_c$. At $|t_1|$ which is not too large we have that
$|{\bf X}(t_1, \bm x)-{\bf X}(t_1, \bm x)|\sim \exp[|\lambda_d t_1|]x$ is much smaller than $\eta$, so that $\omega[{\bf X}(t_1, {\bf x})]\approx \omega[{\bf X}(t_1, 0)]$ and we obtain $F(t_1)\approx E(0)$.
We have
\begin{eqnarray}&&
F(t, \bm x)\approx E(0),\ \ t\ll \frac{1}{|\lambda_d|}\ln \left(\frac{\eta}{x}\right).
\end{eqnarray}
It follows that the divergent part of $g(\bm x)$ is
\begin{eqnarray}&&
g(\bm x)=\frac{\epsilon^2 E(0)}{|\lambda_d|}\ln \left(\frac{\eta}{x}\right)+\ldots,
\end{eqnarray}
where the dots stand for the terms that vanish in the limit $\epsilon\to 0$ uniformly. These terms are small for small $\epsilon$ for all $x$,
in contrast to the first term which is not small for sufficiently small $x$ for any $\epsilon$. The fact that $\eta$ is only determined up
to a constant $C$ of order unity is of no consequence for the answer as becomes clear from the following. The resulting correlation function of $n_s$ is given by
\begin{eqnarray}&&
\langle n_s(0)n_s(\bm x)\rangle=\left(\frac{\eta}{x}\right)^{2C_{KY}}, \label{answer}
\end{eqnarray}
where $C_{KY}$ is the Kaplan-Yorke codimension introduced before. We observe that if we multiply $\eta$ by $C$ then the answer is multiplied by
$C^{2C_{KY}}$ which is close to unity by $C_{KY}\ll 1$.

Eq.~(\ref{answer}) shows that at weak dissipation the correlation codimension is twice the Kaplan-Yorke codimension. Note that the
statistics is isotropic at small $\bm x$ independently of isotropy of $\bm V$. The origin of this isotropy is that the separation vector between two infinitesimally close trajectories grows with an exponent independent of its initial direction.

\subsection{Absence of fluctuations beyond a vanishingly small scale}

Here we notice a very important consequence of the results of the previous subsection. The answer (\ref{structure}) clearly can be written as
\begin{eqnarray}&&
\langle n_s(0)n_s(\bm x)\rangle\approx \exp\left[\epsilon^2 h(\bm x)\right],
\end{eqnarray}
where $h(\bm x)$ is a function that is independent of $\epsilon$ and finite everywhere except for $\bm x=0$. It follows that one can introduce
a scale $l(\epsilon)$ such that
\begin{eqnarray}&&
\langle n_s(0)n_s(\bm x)\rangle\approx 1,\ \ x\gg l(\epsilon),\ \ \lim_{\epsilon\to 0} l(\epsilon)=0.
\end{eqnarray}
The scale $l(\epsilon)$ is not a scale defined up to a factor of order unity, but rather it is a whole range of scales, as
can be easily inferred from Eq.~(\ref{answer}) that gives
\begin{eqnarray}&&
\langle n_s(0)n_s(\bm x)\rangle\approx 1,\ \ 2C_{KY}\ln\left(\frac{\eta}{x}\right)\ll 1.
\end{eqnarray}
For example the scale $l(\delta)$ such that $\langle n_s(0)n_s(\bm x)\rangle-1=\delta\ll 1$ is given by
\begin{eqnarray}&&
2C_{KY}\ln\left(\frac{\eta}{l(\delta)}\right)\approx \delta, \ \ l(\delta)=\eta\exp\left[-\frac{\delta}{2C_{KY}}\right],
\end{eqnarray}
so that
\begin{eqnarray}&&
l(\epsilon^{\beta})=\eta\exp\left[-\frac{|\lambda_d|\epsilon^{\beta-2}}{2E(0)}\right],
\end{eqnarray}
and we get a whole range of scales with $0<\beta<2$ where the transition to $\langle n_s(0)n_s(\bm x)\rangle\approx 1$ takes place. This
situation occurs because we deal with the range in which a power-law holds so there is no possibility of a proper definition of
the correlation length inside that range. Nevertheless the fact is that the scale beyond which the fluctuations of density are weak vanishes
with $\epsilon$. Thus the attractor is almost space-filling and, after a coarse-graining over a scale that vanishes with the dissipation, the attractor fills the whole space. This fact is at the basis of the analysis in the following subsection.

\subsection{Statistics of mass in a small ball}

As it is clear from the explanation of the notion of the weak solution above, the description of $n_s$ via smooth functions is realized by studying
the behavior of $m_l(\bm x)$ at small $l$. We will study the coarse-grained density field $n_l(\bm x)=m_l(\bm x)/V_l$. The root mean square deviation of this field can be easily inferred from the expression for the pair-correlation function derived before. As it is clear from the previous analysis $n_l(\bm x)\approx 1$ for $l$ much larger than a scale that vanishes with $\epsilon$.

The integer moments of $n_l$ or $m_l$ can be inferred from the expressions for the correlation functions derived before. Here we derive
the whole statistics which we also give insight into the build-up of singular density.

We consider the mass $m_l(t=0, \bm x)$ by tracking its pre-image backward in time. The ball is transformed into an ellipsoid centered at
$\bm q(-t, \bm x)$ which largest axis is given by $l\exp[\rho_d(-t)]$ with $\lim_{t\to\infty}\rho_d(-t)/t=|\lambda_d|$. We introduce a
separation scale $L$ which is much smaller than the scale of variations of the smooth velocity field $\bm V$, but over which the density self-averages and its fluctuations are small. It is here that we make the crucial use of the weak dissipation because such a separation scale $L$ exists only
in the regime of weak dissipation. Now we introduce there is a time $t_*$ when the length $l \exp[\rho_d(-t_*)]$ of the largest axis of the ellipsoid
becomes equal to $L$ for the first time.
At that moment of time the mass is spread over a region where the density self-averages and thus the mass
equals to just the volume of that region, $V_l\exp[\rho(-t_*, \bm r)]$. Dividing by $V_l$ we find
\begin{eqnarray}&&
n_l(\bm r)=\exp[\rho(-t_*, \bm r)].
\end{eqnarray}
The probability density function (PDF) of $t_*$ is peaked strongly at $t_*=\ln(L/l)/|\lambda_d|$. Since $\rho(t, \bm r)$
is proportional to $\epsilon$, then the moments of $n_l(\bm r)$ of not too high order are determined by the bulk of the PDF of $t_*$ with corrections
including $\epsilon$. The proof uses the large deviations form of the joint PDF of $\rho(-t)$ and $\rho_d(-t)$
\begin{eqnarray}&&
P(\rho, \rho_d, -t)\propto \exp\left[-tS\left(\frac{\rho}{t}, \frac{\rho_d}{t}\right)\right],
\end{eqnarray}
where $S(x_1, x_2)$ is the rate function, or entropy function, see \cite{review}) references therein. The average
\begin{eqnarray}&&
\langle \exp\left[\alpha\rho\right]\rangle\propto \int d\rho d\rho_d \exp\left[\alpha\rho-t S\left(\frac{\rho}{t}, \frac{\rho_d}{t}\right)\right]
\end{eqnarray}
is found by the saddle-point method. One finds that the saddle-point value of $\rho_d$ is equal to $\gamma t$ where $\gamma=|\lambda_d|+o(\epsilon)$. Thus the moments of the volume $\exp[\rho(-t_*, \bm r)]$ are determined approximately by the most probable events for $\rho_d$ i. e. by $t_*=\ln(L/l)/|\lambda_d|$. We conclude that
\begin{eqnarray}&&
n_l(\bm r)=\exp\left[\rho\left(-\frac{1}{|\lambda_d|}\ln\left(\frac{L}{l}\right), \bm r\right)\right].\label{a4}
\end{eqnarray}
This is a fundamental formula of the weakly dissipative regime: the density fluctuates at the scale $l$ due to mass condensation from
a volume with size smaller than $\eta$ over which the mass is effectively distributed uniformly. This formula allows to find the complete
statistics. The sum rule (\ref{s1}) implies that to order $\epsilon^2$ one must have the equality $-\langle \rho^2(-t)\rangle_c/2=\langle \rho(-t)\rangle$, where $\langle \rho(-t)\rangle=t\sum \lambda_i^-$. Applying the cumulant expansion one finds
\begin{eqnarray}&&
\ln \langle \exp\left[\gamma\rho\left(-t, \bm r\right)\right]\rangle\approx \frac{\gamma^2\langle \rho^2(-t)\rangle_c}{2}+\gamma\langle \rho(-t)\rangle\nonumber\\&&
=-\gamma(\gamma-1)\langle \rho(-t)\rangle=\gamma(\gamma-1)\epsilon^2 E(0)t/2
\end{eqnarray}
Using Eq.~(\ref{a4}) and the absence of fluctuations at the scale $L$ (implying $(\eta/L)^{2C_{KY}}\approx 1$) we find that $n_l$ has
log-normal statistics
\begin{eqnarray}&&
\langle n_l^{\gamma}\rangle=\left(\frac{\eta}{l}\right)^{C_{KY}\gamma(\gamma-1)},
\end{eqnarray}
cf. Eq.~(\ref{bueno}). For $\gamma=2$ the above formula reproduces the answer that the correlation codimension is equal to twice the Kaplan-Yorke
dimension, that was obtained heuristically before.

The spectrum of the fractal dimensions $D(\alpha)$ \cite{HentshelProccacia,BecGawedzkiHorvai},
\begin{eqnarray}&&
D(\alpha)\equiv\lim_{l\to 0}\frac{\ln \langle m_l^{\alpha-1}n_{s}\rangle}{[(\alpha-1)\ln l]}. \label{dimensions}
\end{eqnarray}
involves the average with the stationary measure $n_s$, rather than with the Lebesgue measure. To find this average
consider
$\langle n_l^{\alpha-1}n_{s}\rangle=\lim_{T\to\infty}\!
\langle \exp[\alpha\rho(-t_*, \bm r)\!+\!\int_{-T}^{-t_*}\!\!\omega[t, \bm q(t, \bm r)]dt]\rangle$.
For $t_*\gg \tau_c$, using $\rho(-t_*, \bm r)\approx \rho(-t_*+\tau_c, \bm r)$  one
can make independent averaging $\langle \exp\left[\alpha\rho(-t_*, \bm r)+\int_{-T}^{-t_*}\omega(\bm q(t, \bm r)\right]\rangle=\langle
\exp\left[\alpha\rho(-t, \bm r)\right]\rangle\langle \exp\left[\int_{-T}^{-t}\omega(\bm q(t, \bm r)\right]\rangle=\langle n_l^{\alpha}\rangle$,
where we used volume conservation. 
Substituting the answer for $\langle n_l^{\alpha}\rangle$ we find
\begin{eqnarray}&&
D(\alpha)=d-C_{KY}\alpha. \label{dimensions222}
\end{eqnarray}
The general result of \cite{BecGawedzkiHorvai} for $D(\alpha)$ in the model of white noise velocity field in $d=2$ reduces to the above result at small compressibility.

We observe that the fractal dimensions are close to $d$ (for $\alpha\gg 1$ and high moments the lognormal approximation is not valid generally).
The fractal dimension of the attractor $D(0)$ coincides with the space dimension $d$, while the information dimension $D(1)$ is equal to the Kaplan-Yorke dimension.

\section{Summary}

We showed that the stationary measure of the dissipative dynamical system allows a detailed construction and description in the limit of
weak dissipation. Taking into account that generally the properties of the stationary measure are hard to determine, this solvable
case seems to present significant interest. We showed that the statistics of the stationary measure is log-normal. We derived both the
correlation functions and the statistics of the mass in a small ball. The multi-fractal dimensions are determined completely in this case
by the Kaplan-Yorke dimension that gives the slope of the codimension function $D(\alpha)-d$. The dimensions are close to the full space
dimension with the fractal dimension of the attractor being equal to the space dimension. The information dimension was found equal to the Kaplan-Yorke dimension.

The results are quite general and one can expect that they will find applications. Here we mention as an application the study of the
distribution of inertial particles immersed in the fluid which flow is turbulent, see e. g. \cite{Nature}. In the limit of fast relaxation of the particles' velocity to the speed of the surrounding flow, the particle almost follows the flow. However, this correction is important since
in contrast to the incompressible background flow, the correction brings a finite, albeit small, compressibility to the velocity field. This
is just the case considered in this paper, where the attractor builds up directly in the physical space. The application of the results of
this paper to inertial particles in turbulence will be published elsewhere \cite{IF}. The search for further applications is the subject of
future work.

\section{Acknowledgement}

The author is grateful to J. Yorke for discussions.
This work was supported by COST Action MP$0806$.

\end{document}